C:/comets/C2011L4/SLC L4+S1/130218

# "Secular Light Curves of comets C/2011 L4 Panstarrs and C/2012 S1 ISON Compared to 1P/Halley"

Ignacio Ferrín,

Institute of Physics,

Faculty of Exact and Natural Sciences,

University of Antioquia,

Medellín, Colombia, 05001000

respuestas2013@gmail.com

Submitted for publication

**Abstract.** We used 1329 observations of comet C/2011 L4 Panstarrs and 1897 observations of C/2012 S1 ISON to create the Secular Light Curves (SLCs), and to compare them to comet 1P/Halley (3172 observations), the famous comet of 1986. These comets are at the present time approaching perihelion. Thus the information contained in this report is preliminary but serves for planning purposes. The scientific results found in this investigation are:

**(1)** Comets C/2011 L4 and C/2012 S1 turned on farther away than comet 1P/Halley at R= - 6.2±0.1 AU. Since water ice can not sublimate at distances R< -6 AU, these comets must be rich in substances more volatile than water, like CO or $CO_2$.

**(2) C/2011 L4** exhibits a *"Slowdown Distance"*, **SD**, at which the brightness increase rate slows down to a more relaxed pace. The brightness law brakes. This is reminiscent of the same process for comet 1P/Halley and 11 other comets listed in Table 1. We find R(SD) = -3.6±0.1 AU, m(SD)= 7.1±0.1, while for comet 1P/Halley R(SD) = -1.7±0.1 AU, m(SD)= 5.6±0.1.

**(3)** The absolute magnitude of comet **C/2011 L4** is m(Δ,-R) = m(1,-1)=+5.2 compared with m(1,-1)= +3.9 for comet 1P/Halley. After passing the SD, the comet is increasing its brightness with a shallow power law $R^{+1.25}$. We predict a magnitude at perihelion m(q)= -2.4±0.4, but the comet will be in the Sun's glare.

**(4) Comet C/2012 S1 ISON** poses a dilema about its future luminosity behavior: (A) On the one hand Figure 5 shows evidence of a future event of SD. In this case the future brightness is uncertain and we have to wait until the SD event happens. We assign a probability of 25% to this possibility. (B) On the other hand, Figure 4 shows evidence that this comet *had a recent "Slowdown Distance Event"*, maybe at R(SD) = -7.2±0.1 AU, m(SD)= +12.2±0.1. In this case the power law keeps its pace, and then the comet will be at magnitude m(1,q) = -16±2 at perihelion, brighter than the full moon. Because the orbit enters the Roche Limit, and the calculated temperature at perihelion is 2919° K, the comet may desintegrate giving an awesome cosmic spectacle. We assign to this possibility a probability of 75%.

**(5)** The photometric age, P-AGE, of the comets has been estimated. We find that for both comets P-AGE < 3 comet years which corresponds to "baby comets".

**(6)** The m(SD) vs R(SD) plot shows the Jupiter family comets concentrated in a small area of this phase space. The reason for this is not understood. Some plots shown in this work exhibit complexity beyond current understanding.

## 1. Introduction

2013 offers the opportunity to observe two new comets coming from the Oort cloud, C/2011 L4 Panstarrs and C/2012 S1 ISON. Both have orbits with excentricity e = 1.0.

In June 6$^{th}$, 2011, astronomers at the Institute of Astronomy of the University of Hawaii, discovered a comet with the designation C/2011 L4 Panstarrs (Wainscoat and Micheli, 2011).

We will show that when the comet reached to R= -3.6 AU (the minus sign indicates pre-perihelion location), the brightness law slowed down, from $R^{+8.9}$ to $R^{+1.25}$. We call this the Slowdown Distance, SD. This is not surprising. 12 other comets, including comet 1P/Halley, have exhibited such a SD (Ferrín, 2010) and they are listed in Table 1. The coordinates of the SD are R(SD)= -3.6±0.1 AU, magnitud m(SD)= +7.1±0.1 (see Figure 1).

In September 21$^{st}$ of 2012, Nensky and Novichonok discovered comet C/2012 S1 ISON at the notable distance of -6.3 AU (Nevsky and Novichonok, 2012). The object is coming from the Oort cloud with an excentricity of 1.0000013, esentially parabolic. It has been increasing in brightness at a rate $R^{+4.35}$ and if it were to continue at this rate, it would certainly attain a magnitude much brighter than the full moon. However we will show that this possibility has only a 75% chance of being correct. There is a 25% probability that the comet will exhibit a SD event and attain a much smaller brightness (see Figures 2-5).

In the phase space m(SD) vs R(SD) (see Figure 5) 10 comets lie in a rather small area of the phase space, and two comets C/1995 O1 Hale-Bopp, and C/2011 L4 Panstarrs, lie at the center and left of the diagram. There is no physical explanation of why so many comets lie in such a narrow area of this phase space or why there should a SD. However, a hypothesis is advanced in the text.

## 2. Data sets

The data sets to create the secular light curves are available in internet:

(1) The Cometary Science Archive http://www.csc.eps.harvard.edu/index.html is a site to visit because it contains useful scientific information of current and past comets.
(2) Another usefull site is the Minor Planet Center repository of astrometric observations, http://www.minorplanetcenter.net/db_search .
(3) Seiichi Yoshida's web place http://www.aerith.net contains many raw light curves.
(4) The Yahoo site http://tech.groups.yahoo.com/group/CometObs/ contains up to date observations by many observers for many comets.
(5) The spanish group measures magnitudes with several CCD apertures: http://www.astrosurf.com/cometas-obs. This allows the determination of

   Infinite aperture magnitudes (Ferrín, 2005b), which better represent the whole magnitude of a comet.
(6) The site http://www.observatorij.org/cobs/ contains many observations in the ICQ format, as well as news concerning comets, and it is mantained by the Crni Vrh Observatory.
(7) The german comet group publishes their observations at http://kometen.fg-vds.de/fgk_hpe.htm .

In all we used 1329 observations of comet C/2011 L4 Panstarrs and 1897 observations of C/2012 S1 ISON to create the Secular Light Curves (SLCs), and compared with 3172 observations of comet 1P/Halley.

### 3. SLCs

The elaboration of the Secular Light Curves (SLCs) follows closely the procedures described in the *"Atlas of Secular Light Curves of Comets"* (Ferrín, 2010). The preferred phase space to describe the luminous behaviour is the $m(1,R) = m(\Delta,R) - 5 \log \Delta$ vs Log R plane, where $\Delta$ is the Sun-Earth distance and R the heliocentric distance. In the $m(1,R)$ vs Log R diagram, powers of R, $R^{+n}$, plot as straight lines of slope 2.5 n. The $R^{+n}$ behavior is easy to spot and measure.

In this work we adopted the *"envelope of the data set"* as the correct interpretation of the observed brightness. There are many physical effects that affect comet observations like twilight, moon light, haze, cirrus clouds, dirty optics, lack of dark adaptation, and in the case of CCDs, sky background too bright, insufficient time exposure, insufficient CCD aperture error. All these effects diminish the captured photons emitted by the comet and the observer makes an error downward, toward fainter magnitudes. There are no corresponding physical effects that could *increase the perceived brightness* of a comet. Thus the "envelope" is the correct interpretation of the data. In fact the envelope is rather sharp, while the anti-envelope is diffuse and irregular.

**C/2011 L4 Panstarrs** (Figure 1)**.** To help in the interpretation of the comet's lightcurve, we will compare with the observed SLC of comet 1P/Halley (Ferrín, 2010). The Figure shows that C/2011 L4 Panstarrs turned on much before comet Halley, at -R < -10 AU. Since water can not sublimate at distances beyond -6 AU, the comet must be composed of something more volatile than water, probably CO or $CO_2$. Spectroscopy could tell.

Comet 1P/Halley exhibited a well behaved SLC with a notable *"Slowdown Distance"* after which the rate of brightness increase, diminished. There was a break in the linear law. For 1P the slowdown distance was R= -1.7±0.1 AU. And the deceleration took place at a magnitude of m= +5.6±0.1.

For C/2011 L4 Panstarrs before the SD, the power law was $R^{+8.9}$ and after it is $R^{+1.25}$. Figure 1 gives the absolute magnitude (at $\Delta$=R=1AU) m(1,-1)=+5.2. The magnitude of the comet can then be predicted

$$m(\Delta, R) = m(1,-1) + 5 \log \Delta + 2.5 \, n \log R \qquad (1)$$

This formula predicts a magnitude at perihelion, $m(q) = m(1.109, 0.30) = -2.4 \pm 0.4$. The comet should be as bright at the planet Venus but difficult to see due the the Sun's glare.

**C/2012 S1 ISON** (Figures 2-5). Comparison will also be made to comet 1P/Halley. The comet exhibits a shallow brightness increase with a power law is $R^{+4.35}$ (Figure 2). If this rate were to continue up to perihelion, the comet would indeed be much brighter than the full Moon. This comet poses a dilema:

(A) On one hand, Figure 5 shows evidence of a future event of SD: The comet is approaching the Oort Cloud region of comets. In this case after the SD event, the new power law will allow a definitive prediction of the brigtness of the comet. The predicted location of the SD is shown in Figure 4. The SD event has to take place before October 21$^{st}$, moment at which the comet reaches to 1.25 AU from the Sun, the smallest distance at which a SD event has taken place. For example in that case the comet could pass from a $R^{+4.35}$ power law to the $R^{+1.25}$ very relaxed pace of comet C/2011 L4. This is not very probable, however, because such a shallow power law is very uncommon. Additionally, the two comets are not dynamically related. Thus we assign a probability of 25% to this case.

(B) On the other hand, Figure 4 shows strong evidence that *this comet had a recent "Slowdown Distance Event"*. If we assume the same slope of the SLC as that of comet Halley, then the event could have taken place at $R(SD) = -7.2 \pm 0.1$ AU, $m(SD) = +12.2 \pm 0.1$ (see Figure 2, bottom left corner). Then the current power law is the final law after the SD and it should keeps its pace. A few lines below we calculate that the comet will be at magnitude $m(1,q) = -16 \pm 2$ at perihelion, brighter than the full moon. We assign a probability of 75% to this case.

Using $T = 325°/\sqrt{d}$, $d = q = 0.0124$ AU, we find $T(q) = 2919°$ K. It is questionable if the comet is going to survive such a temperature.

It is possible to calculate the *Roche Limit* for the Sun, using a density of 1 gm/cm$^3$ for the comet. We find $R(Roche) = 1.92 \, 10^6$ km. While the comet reaches perihelion at $q = 0.012472$ AU $= 1.87 \, 10^6$ km. Thus the comet *enters* the Roche Limit of the Sun. The high temperature and the Roche Limit suggest that this comet is going to desintegrate in a cosmic spectacle.

Figure 3 gives the absolute magnitude (at $\Delta = R = 1$ AU) $m(1,-1) = +4.6$. Using $R^{+4.35}$ the magnitude of the comet can then be predicted

$$m(\Delta, R) = m(1,-1) + 5 \log \Delta + 2.5 \, n \log R \qquad (2)$$

$$= 4.6 + 5 \log \Delta + 10.9 \log R$$

This formula predicts a magnitude at perihelion $m(q) = m(0.99, 0.01) =$

-16±2, brighter than the full moon.

## 4. Photometric Age, P-AGE

The photometric age of a comet is a proxy for age and may be defined as (Ferrín, 2010):

$$\text{P-AGE [comet years]} = 1440 / [\, A_{sec}\, R_{sum}\,] \qquad (3)$$

where $A_{sec}$ = amplitude of the secular light curve = $V(1,1,0) - m_1(1,1)$, and $R_{sum}$ is the sum of the turn on, $R_{on}$, and turn off, $R_{off}$, distances of the comet. P-AGE is measured in comet years to be sure that they are not confused with current years. The constant is chosen so that comet 29P/Neujmin 1 has a P-AGE = 100 cy. For both comets in this work we do not know any of these parameters so it would seem impossible to determine P-AGE. However in Figure 6 we show a correlation P-AGE vs $R_{on}$. Since we know that $R_{on} < -10$ AU for both comets, Figure 6 returns P-AGE < 3 cy. In the classification of Ferrín (2010) these are "baby comets" ( those with P-AGE < 4 cy ).

## 5. m(SD) vs R(SD) diagram.

This diagram is shown in Figure 5 and it is based on the data presented in Table 1. It shows 10 comets closely located in a small area of this phase space, and two comets beyond their area. Most of these comets are members of the Jupiter family but notice that 4 members of the Oort Cloud are located inside the Jupiter Family Box. The reason why Oort Cloud and JF comets are segregated in this diagram, is not understood.

## 6. Why should there be a SD?

When a comet from the Oort cloud (e=1.0) falls into the Sun, the upper layer of the nucleus contains fresh volatiles like CO, $CO_2$ an $H_2O$. As the comet approaches, temperature increases, and the first one to sublimate is CO. Next sublimates $CO_2$ and finally $H_2O$. This is due to their vapor pressure. Sublimating CO or $CO_2$, the light curve far from the Sun is a straight line with a power law ~ $R^{+9.1\pm2.0}$ (Table 2). The $H_2O$ does not have sufficient temperature to sublimate and thus CO or $CO_2$ control the surface sublimation and the light curve. The sublimation rate increases as the comet approaches the Sun, and at a given temperature, $H_2O$ overpowers CO or $CO_2$, and $H_2O$ now controls the surface. The brightness increase decreases its rate according to the new sublimation rules. This is a hypothetical mechanism behind the SD. However, there have been no numerical or theoretical models to explain this discontinuity, and thus the hypothesis remains unconfirmed.

R(SD) is a distance and it must be related to a temperature. Observationally (Figure 5) we find a range $1.24 < R(SD) < 2.09$. Using $T = 325°/\sqrt{d}$, we find $222° < T < 289°$ K. This means that the comets in our data base have had their events within a range of temperatures of only 67°K, perhaps due to different pole orientations.

m(SD) would seem to be related to the ratio CO/H2O or CO2/H2O. It is expected that espectroscopic and compositional estudies of C/2011 L4 Panstarrs and C/2012 S1, would reveal differences that can be correlated to the list of SD's properties compiled in Table 1 including JF Comets.

**7. Conclusions**

**(1)** We present the secular light curves, SLCs, of comets C/2011 L4 Panstarrs and C/2012 S1 ISON. Both comets turned on beyond -10 AU from the Sun. For comparison comet 1P/Halley turned on at R= - 6.2±0.1 AU. Since water ice can not sublimate at distances R<-6 AU, these comets have to contain substances more volatile than water, like CO or CO2.

**(2)** We measure the *Slowdown Distance* of C/2011 L4 Panstarrs. This is the distance at which the brightness increase rate slows down to a more relaxed pace. The brightness law brakes. This is reminiscent of the same process for comet 1P/Halley and 11 other comets listed in Table 1. We find R(SD) = -3.6±0.1 AU, m(SD)= 7.1±0.1, while for comet 1P/Halley R(SD) = -1.7±0.1 AU, m(SD)= 5.6±0.1.

**(3)** We derive the absolute magnitude of C/2011 L4 Panstarrs and the power laws that define its brightness behavior. The absolute magnitude is m(Δ,-R) = m(1,-1)=+5.2 compared with m(1,-1)= +3.9 for comet 1P/Halley. After passing the SD, the comet is increasing its brightness with a shallow power law $R^{+1.25}$. With this information the magnitude at perihelion can be calculated and we find m(q)= -2.4±0.4 .

**(4) Comet C/2012 S1 ISON** poses a dilema about its future luminosity behavior: (A) On the one hand Figure 5 shows evidence of a future event of SD. In this case the future brightness is uncertain and we have to wait until the SD event happens. We assign a probability of 25% to this possibility. (B) On the other hand, Figure 4 shows evidence that this comet *had a recent "Slowdown Distance Event",* maybe at R(SD) = -7.2±0.1 AU, m(SD)= +12.2±0.1. In this case the power law keeps its pace, and then the comet will be at magnitude m(1,q) = -16±2 at perihelion, brighter than the full moon. Because the orbit enters the Roche Limit, and the calculated temperature at perihelion is 2919º K, the comet may desintegrate giving an awesome cosmic spectacle. We assign to this possibility a probability of 75%.

**(5)** The photometric age, P-AGE, of the comets has been estimated. We find that for both comets P-AGE < 3 comet years which corresponds to "baby comets".

**(6)** The m(SD) vs R(SD) plot shows the Jupiter family comets concentrated in a small area of this phase space. The reason for this is not understood. Some of the plots shown in this work, exhibit complexity beyond current understanding.

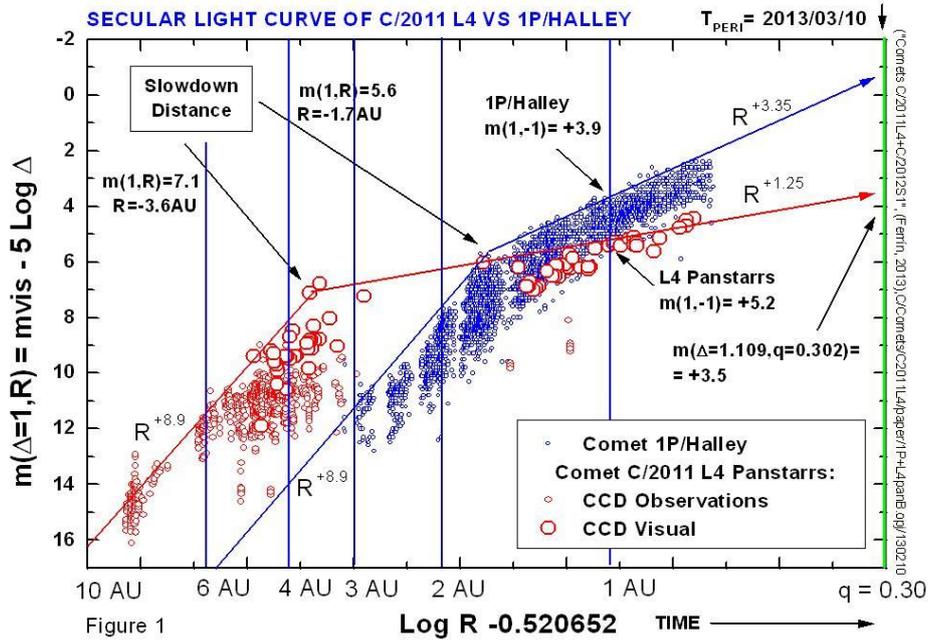

*Figure 1. The secular light curve of comet **C/2011 L4** compared with that of comet 1P/Halley. A number of features can be discerned. The turn on distance of **C/2011 L4** is much farther away than that of 1P, in fact beyond 10 AU. Since water cannot sublimate a R>6 AU the comet has to be made of something more volatile than water, like CO or $CO_2$. Both comets exhibit a "slowdown distance". For 1P it is located at R= -1.7±0.1 AU, while for **C/2011 L4** it is located at R= -3.6±0.1 AU. For comparison comet C/1995 O1 Hale-Bopp had R= -6.4±0.1 AU. The absolute magnitudes of 1P and **C/2011 L4** are m(1,-1)= +3.9 and +5.2, but C/2011 L4 exhibits a very low magnitude increase $R^{+1.25}$ vs comet 1P/Halley $R^{+3.35}$. Curiously both comets exhibit the same slope before SD. The reason why we select the envelope as the correct interpretation of the light curve, is explained in the text. We used 1329 observations of the comet to create the SLC. For comet Halley we used 3172 observations.*

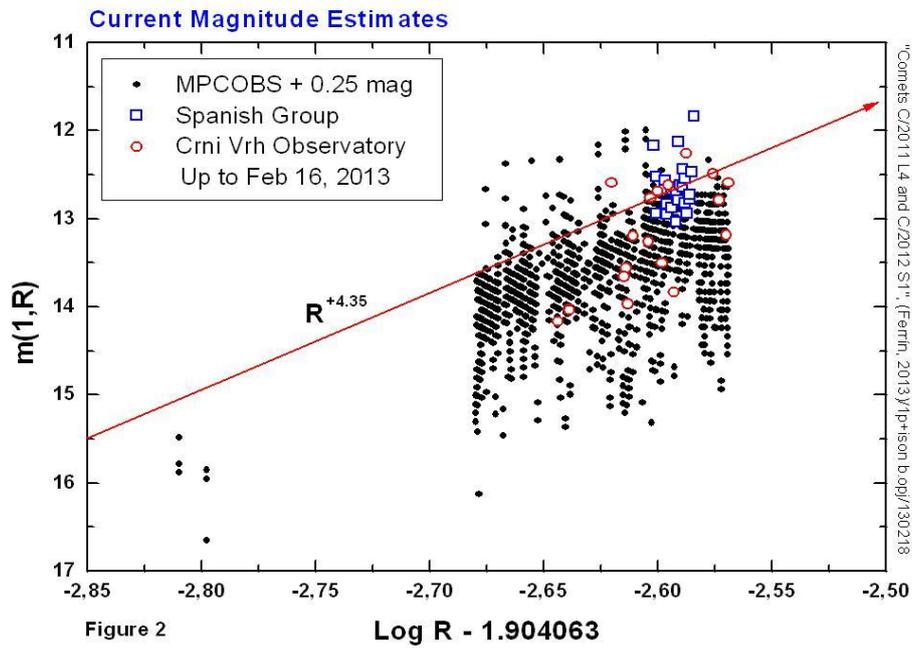

***Figure 2.*** *Enlarged region around the current time location of comet **C/2012 S1** with up to date observations (2013 Feb 16$^{th}$). The brightness is increasing with a power law $R^{+4.35}$ (Table 1). This shallow value makes it probable that the comet has already passed its SD phase. In this case the current rate serves as a predictor. Or the comet will experience a SD event, in which case the rate* after *the event will serve as a predictor. For comet Halley we used 3172 observations. For comet **C/2012 S1** 1897 observations.*

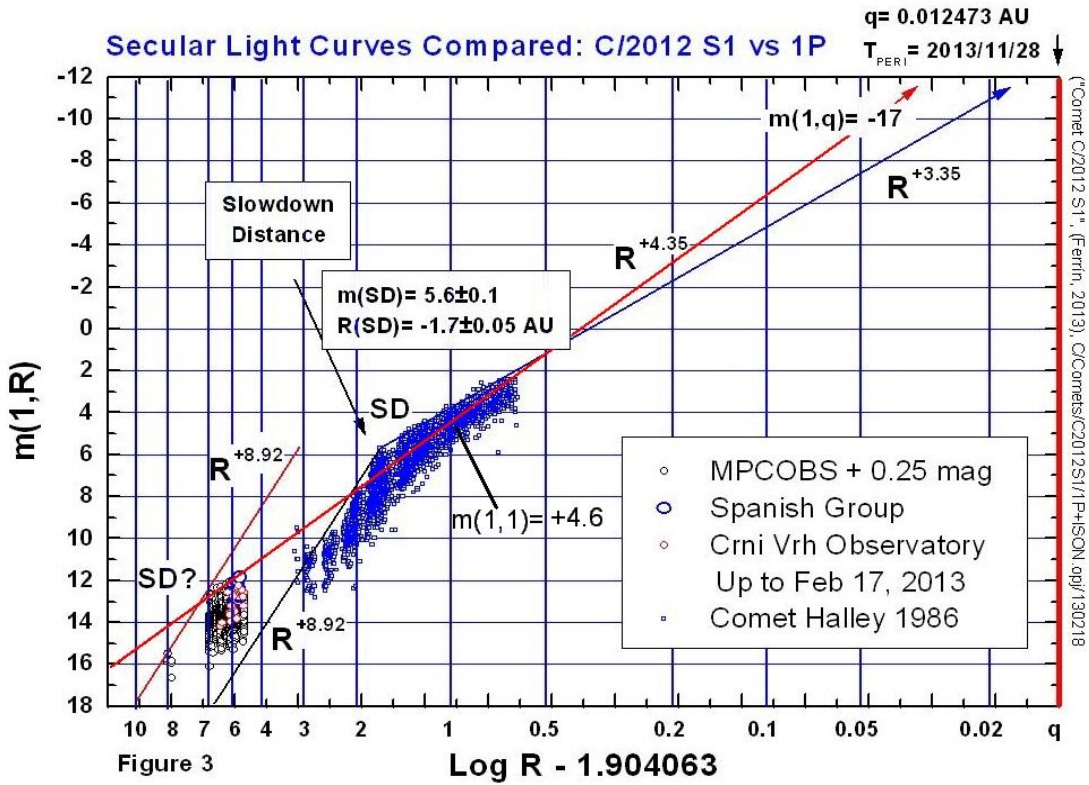

*Figure 3.* The secular light curve of comet **C/2012 S1** compared with that of comet 1P/Halley. A number of features can be discerned. The turn on distance of **C/2012 S1** is much farther away than that of 1P. Since water cannot sublimate a R> -6 AU, the comet has to be composed of something more volatile than water, like CO or $CO_2$. We find the absolute magnitude $m(1,1)= +4.6\pm0.2$. Comet Halley exhibits a "slowdown distance" located at $R(SD)= -1.7\pm0.1$ AU, $m(SD)= +5.6\pm0.1$. For comet C/2012 S1 we have two possibilities:

(A) The comet has not yet arrived to its SD event, as suggested by Figure 5. In which case the final brightness is too early to tell and we have to wait until past the SD event.

(B) If we assume a slope identical to those of comet 1P/Halley and **C/2011 L4**, that is $R^{+8.92}$, then comet **C/2012 S1** may have had a SD event at $R = -7.2\pm0.1$ AU, $m(1,R)= +12.2\pm0.1$ (lower left part of the diagram), in which case the observed power law is the power law after SD event. Figure 4 also suggests very strongly that the shallow power law exhibited by this comet is in accord with that exhibited by Oort Cloud comets, after SD events (confirmation Table 2). In this case the comet would reach to magnitude $m(1,q)=-17$ at perihelion, brighter than the full Moon, if the pace is sustained, as has been sustained in previous cases.

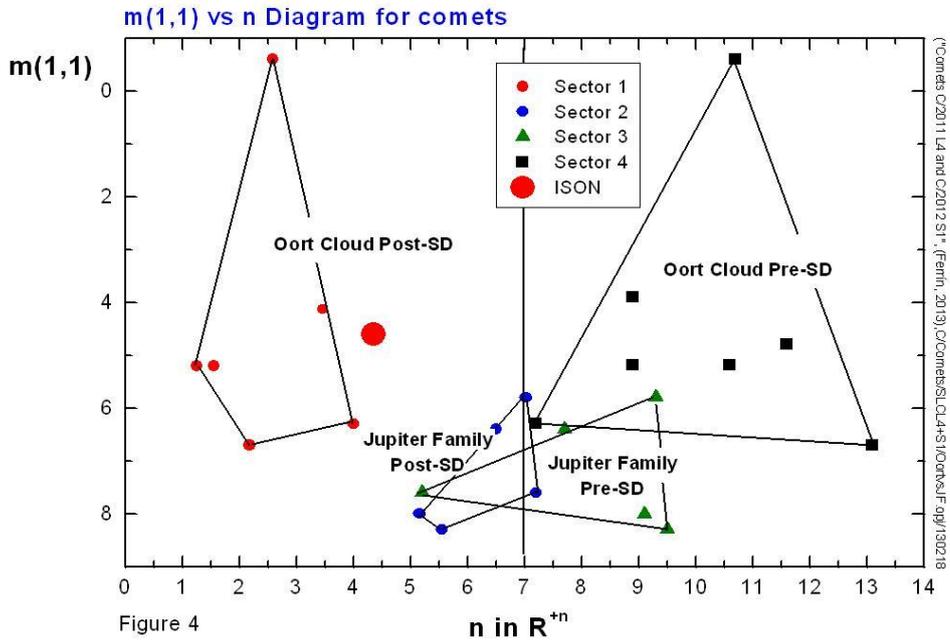

***Figure 4.*** *A m(1,1) vs n Diagram for Comets. Or the Pre-SD vs Post-SD vs Oort Cloud vs JF comets diagram. The values measured in Table 2 are plotted here. They separate Pre and Post-SD events, and Oort Cloud Comets vs Jupiter Family comets into 4 distinct sectors. In particular the Pre-Post SD of Oort Cloud comets are clearly separated. The location of comet C/2012 S1 is nearer to Oort Cloud Comets in a Post-SD Phase. In this case the current power law, $R^{+4.35}$, is a good estimate of its future brightness.*

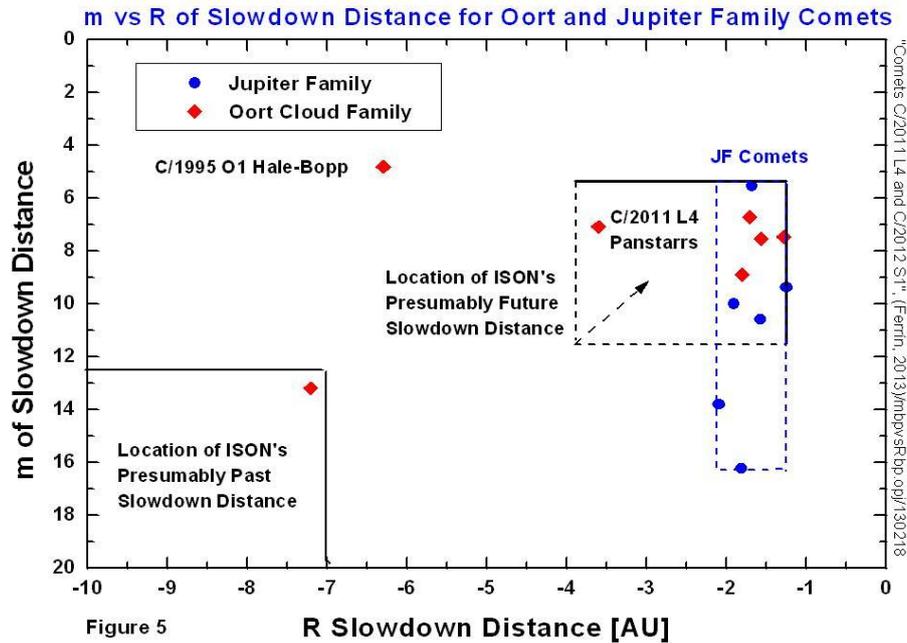

***Figura 5.*** *The Slowdown Magnitudes of 13 comets are plotted vs their Slowdown Distances from Table 1. 10 of 12 (83%) lie in a narrow vertical zone centered at R = -1.7 AU. Another comet (C/2011 L4) lies in the middle of the diagram, and another (C/1995 O1 Hale-Bopp) lies to the left. The location of the Slowdown Distance of comet C/2012 S1 in indicated in two regions: (A) if the event already passed, down and to the left. (B) If the event is going to happen, to the right. Notice that the comet is moving toward the region of other Oort Cloud comets. This favors option (B) (that the SD event has not happened). However Figure 4 favors the opposite, that the event has taken place. Table 1 allows us to define the Jupiter Family Box of Comets, as 1.24 < R(SD) < 2.09 AU, and 4.82 < m(SD) < 13.81. Inside this box there are 4 Oort Cloud comets. Up to this moment there is no physical interpretation of why so many comets should lie in such a small area of this phase space, although a preliminary hyphothesis is advanced in the text.*

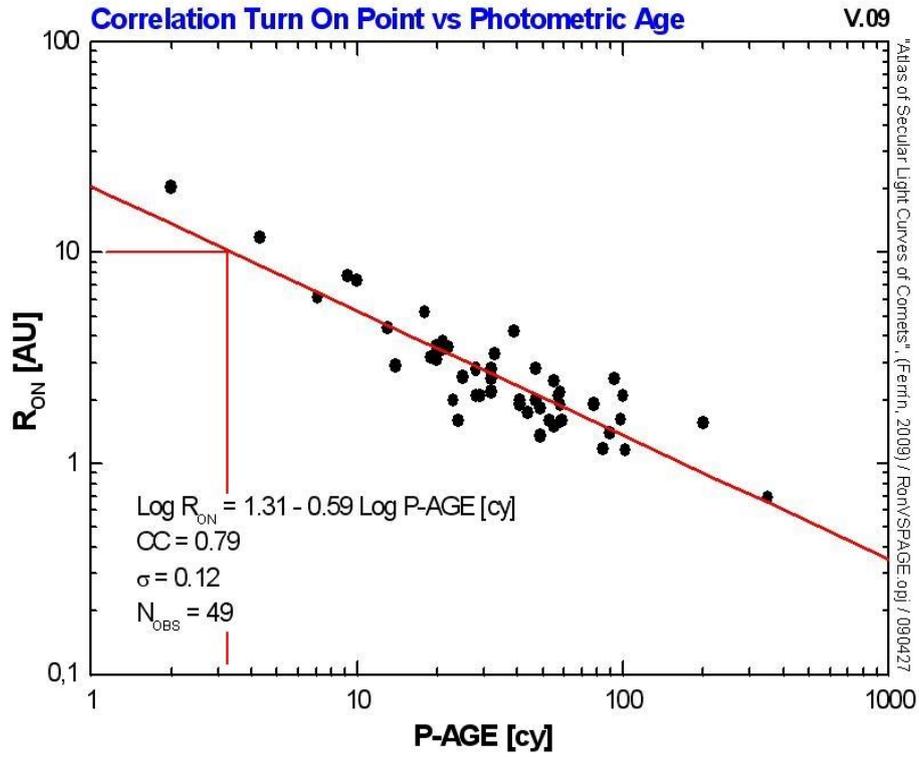

*Figure 6.* The correlation between the turn on point Ron vs the photometric age P-AGE measured in comet years. The correlation shows that comets with turn on points beyond 10 AU are less than 3 comet years old. Older comets have to get nearer to the Sun to get activated. The two comets studied in this work, turned on beyond 10 AU and thus are classified as "baby comets" (Ferrín, 2010).

Table 1.  Distances, magnitudes at Slowdown,q,e,i,Tiss.

```
------------------------------------------------------------
R(SD)  m(SD)   q      e         i    Tiss  Comet
------------------------------------------------------------
-1.8    8.9   0.316  1.000240  119.9  0.00  C/1956 R1 Arend-Roland
-3.6    7.1   0.302  1.000028   84.3  0.00  C/2011 L4 Panstarrs
-1.56   7.55  0.171  1.000018   77.1  0.00  C/2006 P1 McNaught
 ----   ----  0.012  1.000004   62.1  0.00  C/2012 S1 ISON
-1.39   7.6   0.099  0.999903   81.7  0.06  C/2002 V1 NEAT
-1.7    6.73  0.230  0.999758  124.9 -0.33  C/1996 B2 Hyakutake
-6.29   4.82  0.914  0.994929   89.4  0.04  C/1995 O1 Hale-Bopp
-1.68   5.55  0.586  0.967813  162.3 -0.61  1P/Halley
------------------------------------------------------------
-1.57  10.6   1.031  0.707045   31.9  2.46  21P/Giacobinni-Zinner
-1.24   9.38  1.059  0.694533   13.6  2.64  103P/Hartley 2
-1.81  16.24  1.057  0.659295   11.7  2.81  46P/Wirtanen
-1.9   10.0   1.598  0.537385    3.2  2.88  81P/Wild 2
-2.09  13.81  1.509  0.516946   20.5  2.90  9P/Tempel 1
------------------------------------------------------------
```

Table 2. Absolute magnitudes and power laws.

```
Comet                m(1,1)          n in R^n
------------------------------------------------------------
Oort Cloud Comets                 Pre SD     Post SD
------------------------------------------------------------
C/1956 R1 AR         +6.3±0.1      +7.2 and +4.0
C/2011 L4 Pan        +5.2±0.1      +8.9 and +1.25²
C/2006 P1 McN        +5.2±0.1     +10.6 and +1.55
C/2012 S1            ------        ---- ---  +4.3²
C/2002 V1 NEAT       +6.7±0.1     +13.1 and +2.18
C/1996 B2 Hy         +4.8±0.1     +11.6 and +2.33
C/1995 O1 HB         -0.6±0.1     +10.7 and +2.58
1P/Halley            +3.9±0.1      +8.9 and +3.35
------------------------------------------------------------
Jupiter Family
------------------------------------------------------------
21P                  +8.0±0.1      +9.1 and +5.16
103P                 +8.3±0.1      +9.5 and +5.55
46P                  +7.6±0.1      +5.2 and +7.20
81P                  +5.8±0.2      +9.3 and +7.03
9P                   +6.4±0.2      +7.7 and +6.50
------------------------------------------------------------
```
1 Comet C/2011 L4 has the smallest Post SD n value
  of the whole sample.
2 C/2012 S1 has a small rate of brightness increase
  reminiscent of Oort Cloud Comets Post SD.  This
  data is plotted in Figure 4.